\documentclass[a4paper,11pt]{article}

\usepackage{amsmath,amssymb,mathtools}
\allowdisplaybreaks[1]
\usepackage{color}
\usepackage{graphicx}
\usepackage{subfigure}
\usepackage{cite}
\usepackage[colorlinks=true,linkcolor=red,citecolor=blue,urlcolor=blue,bookmarks]{hyperref}
\usepackage{multirow,makecell}
\usepackage{textcomp}
\usepackage{wasysym}
\usepackage{ulem}

\usepackage{verbatim}

\usepackage[utf8]{inputenc}
\usepackage[T1]{fontenc}

\usepackage{array}

\usepackage{diagbox}  

\usepackage[text={17.1cm,24.6cm},centering]{geometry} 

\numberwithin{equation}{section}

\def \be {\begin{equation}}
\def \ee {\end{equation}}
\def \ba {\begin{array}}
\def \ea {\end{array}}
\def \bea {\begin{eqnarray}}
\def \eea {\end{eqnarray}}
\def \nn {\nonumber}

\def \a {\alpha}
\def \b {\beta}

\def \G {\Gamma}

\def \D {\Delta}
\def \e {\epsilon}
\def \ve {\varepsilon}

\def \s {\sigma}

\def \r {\rho}

\def \cA {\mathcal A}
\def \cB {\mathcal B}

\def \cD {\mathcal D}
\def \cE {\mathcal E}
\def \cF {\mathcal F}

\def \cH {\mathcal H}
\def \cI {\mathcal I}

\def \cL {\mathcal L}

\def \cT {\mathcal T}

\def \cX {\mathcal X}
\def \cY {\mathcal Y}

\def \p {\partial}

\def \f {\frac}

\def \td {\tilde}

\def \inf {\infty}

\def \lag {\langle}
\def \rag {\rangle}

\def \ii {\mathrm{i}}

\def \tr {\textrm{tr}}
\def \tdtr {\widetilde{\textrm{tr}}}

\def \and {{~\textrm{and}~}}

\def \CFT {{\textrm{CFT}}}

\def \holo {{\textrm{holo}}}

\def \AdS {{\textrm{AdS}}}
\def \CFT {{\textrm{CFT}}}

\begin{document}

\title{
\textbf{Subsystem quantum R\'enyi-Bhattacharyya distance\\in 2D CFTs and its holographic applications}
}

\author{
Yihao Wang
and
Jiaju Zhang\footnote{Corresponding author: jiajuzhang@tju.edu.cn}
}
\date{}
\maketitle
\vspace{-10mm}
\begin{center}
{\it
Center for Joint Quantum Studies and Department of Physics, School of Science, Tianjin University, 135 Yaguan Road, Tianjin 300350, China
}
\vspace{10mm}
\end{center}

\begin{abstract}

  We develop a systematic method to compute the subsystem fidelity and R\'enyi fidelity in two-dimensional conformal field theories (CFTs) from the operator product expansion of twist operators, and apply it to the subsystem quantum Bhattacharyya distance (BD) and R\'enyi-Bhattacharyya distance (RBD). We validate the method against exact results in the free massless fermion and boson theories, and then use it to compute the BD and RBD in holographic CFTs between two heavy primary states, between two thermal states, and between a heavy primary state and a thermal state. When the two states differ in their classical, order-$c$ data, the distances are of order $c$, and the states are distinguishable already at the level of classical gravity; when the leading data coincide, the distances are of order $1/c$ at every order of the short interval expansion, so that the distinguishability becomes a perturbative quantum gravity effect. For a heavy primary state and its thermal counterpart, the leading contribution to the BD appears at order $\ell^8$ and is of order $c^0$, whereas the RBD is generically of order $c$, in line with the earlier finding that a heavy primary state and a thermal state share the same order-$c$ entanglement entropy but have different order-$c$ R\'enyi entropies in the short interval expansion.

\end{abstract}

\baselineskip 18pt
\thispagestyle{empty}
\newpage

\tableofcontents

\section{Introduction}

Quantifying how different two quantum states are is a fundamental problem that appears in many areas of physics. In quantum information theory, distance measures between density matrices underlie quantum hypothesis testing and state discrimination, and they give operational meaning to the distinguishability of quantum states \cite{Nielsen:2010oan,Hayashi:2017sdr,Watrous:2018rgz}. In quantum many-body physics, fidelity-type distances serve as sharp diagnostics of quantum phase transitions and of the distinguishability of many-body states \cite{Gu:2008Fidelity,deBoer:2020snb}. In quantum field theory, entropic and distance measures constrain the structure of the theory, for instance along the renormalization group flow \cite{Casini:2004bw}. In quantum gravity, whether two states can be distinguished by observations restricted to a subsystem is closely related to the black hole information problem \cite{Hawking:1976ra,Penington:2019npb,Almheiri:2019psf}.

Among the various distance measures, the fidelity of two density matrices $\rho$ and $\sigma$ is defined as \cite{Nielsen:2010oan,Hayashi:2017sdr,Watrous:2018rgz}
\be \label{Fdef}
F(\rho,\sigma) \equiv \tr\sqrt{\sqrt\rho\,\sigma\sqrt\rho}.
\ee
The square roots make a direct evaluation of the fidelity in a conformal field theory (CFT) difficult, and one usually resorts to a replica trick \cite{Lashkari:2014yva,Suzuki:2019xdq,Kirklin:2019ror,Kusuki:2019hcg,Parez:2022sgc,Xiao:2023hnh,Sui:2025qve}.
In this paper we use the equivalent form \cite{Baldwin:2022cjb}
\be \label{Fdef2}
F(\rho,\sigma) = \tr\sqrt{\rho\sigma},
\ee
and calculate the fidelity from the replica trick \cite{Kirklin:2019ror,Parez:2022sgc,Xiao:2023hnh,Sui:2025qve}
\be \label{replicatrick}
F(\rho,\sigma) = \lim_{p\to\f12} \tr[(\rho\sigma)^p].
\ee
The R\'enyi fidelity is defined as \cite{Wang:2008gcy,Parez:2022sgc}
\be \label{RFdef}
F_p(\rho,\sigma) \equiv \frac{\tr[(\rho\sigma)^p]}{\sqrt{\tr(\rho^{2p})\tr(\sigma^{2p})}},
\ee
and it has been used in studies of non-equilibrium evolution after quantum quenches \cite{Cardy:2014rqa,Parez:2025wtd,Parez:2025psa} and in bulk reconstruction in the AdS/CFT correspondence \cite{Suzuki:2019xdq,Kusuki:2019hcg}.

For two classical probability distributions $\mathfrak{p}$ and $\mathfrak{q}$, the Bhattacharyya distance is defined as \cite{Bhattacharyya:1943qzv}
\be
B(\mathfrak{p},\mathfrak{q}) = - \log \big( \sum_i \sqrt{\mathfrak{p}_i \mathfrak{q}_i} \big).
\ee
Noting the forms of the fidelity $F(\r,\s)$ \eqref{Fdef2} and the R\'enyi fidelity $F_p(\r,\s)$ \eqref{RFdef} between two quantum states $\r$ and $\s$, one can define the quantum Bhattacharyya distance (BD) and the quantum R\'enyi-Bhattacharyya distance (RBD) as
\be \label{BD}
B(\r,\s) = - \log F(\r,\s),
\ee
\be \label{RBD}
B_p(\r,\s) = - \log F_p(\r,\s).
\ee
The RBD is non-negative, $B_p(\r,\s)\geq0$, and $B_p(\r,\s)=0$ only if $\r=\s$; moreover, $B(\r,\s)=B_{1/2}(\r,\s)$, so that the BD is a special case of the RBD. Neither the BD nor the RBD is a strict metric, since the triangle inequality is not satisfied in general.
The BD and the RBD quantify how well two quantum states can be distinguished from each other, and for reduced density matrices they measure how well two global states can be distinguished by observations restricted to a subsystem. Various measures of this kind have been investigated in 2D CFTs, including the subsystem trace distance \cite{Zhang:2019wqo,Zhang:2019itb}, the subsystem fidelity \cite{Lashkari:2014yva,Zhang:2019itb,Xiao:2023hnh,Zhang:2025vaq,Sui:2025qve}, the subsystem relative entropy \cite{Lashkari:2014yva}, and the subsystem Schatten distances \cite{Basu:2017kzo,He:2017txy}.

In this paper we extend the twist operator OPE method \cite{Headrick:2010zt,Calabrese:2010he,Rajabpour:2011pt,Chen:2013kpa,%
Chen:2013dxa,Chen:2016lbu,Lin:2016dxa,He:2017vyf,Basu:2017kzo,He:2017txy,Ruggiero:2018hyl,Guo:2018pvi,Zhang:2019wqo,Zhang:2019itb} from the fidelity \cite{Zhang:2025vaq,Sui:2025qve} to the R\'enyi fidelity, and use the method to compute the subsystem BD and RBD in 2D CFTs. After validating the method in the free massless fermion and boson theories, we apply it to holographic CFTs and compute the distances between two heavy primary states, between two thermal states, and between a heavy primary state and a thermal state. We find a uniform pattern that has a good explanation in the $\AdS/\CFT$ correspondence \cite{Maldacena:1997re,Gubser:1998bc,Witten:1998qj}. A 2D holographic CFT has a large central charge $c=\frac{3R}{2G}$ \cite{Brown:1986nw}, with $R$ being the AdS radius and $G$ being the Newton constant, and a sparse spectrum of low-lying operators \cite{Hartman:2014oaa}. When the two states differ in their classical, order-$c$ data, the distances are of order $c$, and the states are distinguishable already in classical gravity; when the leading data coincide, the distances are of order $1/c$ at every order of the short interval expansion, and the distinguishability is purely a perturbative quantum gravity effect. For a heavy primary state and its thermal counterpart, the BD starts at order $\ell^8$ of the short interval expansion and is of order $c^0$, whereas the RBD is generically of order $c$, with the classical contribution cancelled only in the fidelity limit $p\to\f12$.

The paper is organized as follows. In section~\ref{secMethod} we present the method for calculating the subsystem fidelity and R\'enyi fidelity from the OPE of twist operators. In sections~\ref{secFermion} and \ref{secBoson} we check the method against exact results in the 2D free massless fermion and boson theories, respectively. In section~\ref{secHoloCFT} we apply the method to 2D holographic CFTs. We conclude in section~\ref{secConc}. The derivation of the universal contribution $\cF^{(1/2)}_{\cX\cX\cY}$ is given in appendix~\ref{appF12XXY}.

\section{Fidelity and R\'enyi fidelity from OPE of twist operators} \label{secMethod}

In this section, we present our method for calculating the subsystem fidelity and the R\'enyi fidelity in 2D CFTs. We first review the calculation of the fidelity from the OPE of twist operators in Ref.~\cite{Sui:2025qve}, collect the universal contributions from several quasiprimary operators, and derive the universal contribution of a new combination of quasiprimary operators. We then present a method for calculating the R\'enyi fidelity using the OPE of twist operators.

\subsection{OPE of twist operators}

We consider a 2D CFT on a cylinder of spatial period $L$. We take an interval $A=[0,\ell]$ and denote its complement by $\bar A=[\ell,L]$. Only translationally invariant states are considered in this paper.

For each translationally invariant state $\r_i$, with the index $i$ labelling different states, we define the reduced density matrix (RDM) $\r_{A,i} = \tr_{\bar A} \r_i$, where the trace is taken over the complement of $A$. For $n$ copies of the RDMs $\r_{A,i}$ with $i=0,1,\cdots,n-1$, which are not necessarily identical, the general ansatz is that
\be
\tr_A( \r_{A,0} \r_{A,1} \cdots \r_{A,n-1} ) = \lag \cT(\ell,\ell) \td \cT(0,0) \rag_{\r_0\otimes\r_1\otimes\cdots\otimes\r_{n-1}},
\ee
where on the right-hand side the expectation value of the twist operators \cite{Calabrese:2004eu,Cardy:2007mb,Calabrese:2009qy} $\cT$ and $\td\cT$, inserted at the two endpoints of the interval, is taken in the direct product state $\r_0\otimes\r_1\otimes\cdots\otimes\r_{n-1}$ of the $n$-fold CFT, denoted as CFT$^n$. The twist operators $\cT$ and $\td\cT$ are primary operators with conformal weights and scaling dimension \cite{Calabrese:2004eu}
\be
h_n = \bar h_n = \f{c(n^2-1)}{24n}, ~~
\D_n = h_n + \bar h_n = \f{c(n^2-1)}{12n}.
\ee

For later convenience, we define the normalized trace as
\be \label{tdtr}
\tdtr_A( \r_{A,0} \r_{A,1} \cdots \r_{A,n-1} ) \equiv \f{1}{c_n} \Big(\f{\ell}{\e}\Big)^{2\D_n} \tr_A( \r_{A,0} \r_{A,1} \cdots \r_{A,n-1} ),
\ee
where $\e$ is a UV cutoff and $c_n$ is a constant related to the normalization of the twist operators. With this normalization, the normalized trace approaches unity as the interval shrinks,
\be
\lim_{\ell \to 0} \tdtr_A( \r_{A,0} \r_{A,1} \cdots \r_{A,n-1} ) = 1.
\ee

Given that all the relevant states are translationally invariant, the OPE of twist operators \cite{Headrick:2010zt,Calabrese:2010he,Rajabpour:2011pt,Chen:2013kpa,Chen:2013dxa,Chen:2016lbu,Lin:2016dxa,He:2017vyf,Basu:2017kzo,%
He:2017txy,Ruggiero:2018hyl,Guo:2018pvi,Zhang:2019wqo,Zhang:2019itb} gives
\be
\tdtr_A( \r_{A,0} \r_{A,1} \cdots \r_{A,n-1} ) = 1+\sum_K \ell^{\D_K} d_K \lag \Phi_K
                                                 \rag_{\r_0\otimes\r_1\otimes\cdots\otimes\r_{n-1}},
\ee
where $\Phi_K$ is a non-identity quasiprimary operator of CFT$^n$, $\D_K$ is its scaling dimension, and $d_K$ is the corresponding OPE coefficient.
Again due to the translational invariance of the relevant states, we only need to include the CFT$^n$ quasiprimary operators $\Phi_K$ that take the form of direct products
\be \label{PhiK}
\Phi_K^{j_1 j_2 \cdots j_k} = \cX_1^{j_1} \cX_2^{j_2}\cdots \cX_k^{j_k}, ~~
\D_K = \D_{\cX_1} + \D_{\cX_2} + \cdots +\D_{\cX_k},
\ee
where $j_1,j_2,\cdots,j_k \in \{0,1,\cdots,n-1\}$ are replica indices and $\cX_1,\cX_2,\cdots,\cX_k$ are quasiprimary operators of a single copy of the CFT with respective scaling dimensions $\D_{\cX_1},\D_{\cX_2},\cdots,\D_{\cX_k}$. The corresponding OPE coefficient $d_{\cX_1\cX_2\cdots\cX_k}^{j_1j_2\cdots j_k}$ can be calculated following \cite{Calabrese:2010he}. We further get
\be \label{general}
\tdtr_A( \r_{A,0} \r_{A,1} \cdots \r_{A,n-1} ) = 1+ \sum_{k=1}^n
                                                    \sum_{\{\cX_1,\cdots,\cX_k\}} \ell^{\D_{\cX_1} + \cdots +\D_{\cX_k}}
                                                    \sum_{j_1,\cdots,j_k} d_{\cX_1\cdots\cX_k}^{j_1\cdots j_k}
                                                    \lag \cX_1 \rag_{\r_{j_1}} \cdots \lag \cX_k \rag_{\r_{j_k}}.
\ee
In the sums over the replica indices $j_1,\cdots,j_k \in \{0,1,\cdots,n-1\}$, one has to be careful about the range of each index, so as to avoid under-counting or over-counting of the CFT$^n$ quasiprimary operators.

When all the copies are identical, $\r_{A,j}=\r_A$, eq.~\eqref{general} gives the moment \cite{Chen:2016lbu,Lin:2016dxa,He:2017vyf}
\be \label{tdtrArAn}
\tdtr_A\r_{A}^n = 1+ \sum_{k=1}^n
                  \sum_{\{\cX_1,\cdots,\cX_k\}} \ell^{\D_{\cX_1} + \cdots +\D_{\cX_k}}
                  b_{\cX_1\cdots\cX_k}
                  \lag \cX_1 \rag_{\r} \cdots \lag \cX_k \rag_{\r},
\ee
and thus the R\'enyi entropy. Here we have defined the coefficient
\be
b_{\cX_1\cdots\cX_k} \equiv \sum_{j_1,\cdots,j_k} d_{\cX_1\cdots\cX_k}^{j_1\cdots j_k}.
\ee
When there is an even number $n=2p$ of alternating states, i.e.~$\r_{A,a}=\r_A$ for even $a=0,2,\cdots,2p-2$ and $\r_{A,b}=\s_A$ for odd $b=1,3,\cdots,2p-1$, we get \cite{Sui:2025qve}
\be \label{tdtrArAsAp}
\tdtr_A[( \r_A \s_A )^p] = 1 + \sum_{k=1}^{2p}
                               \sum_{\{\cX_1,\cdots,\cX_k\}}
                               \ell^{\D_{\cX_1}+\cdots+\D_{\cX_k}}
                               \cF_{\cX_1 \cdots \cX_k}^{(p)},
\ee
with the definition
\be \label{FpX1dddXk}
\cF_{\cX_1 \cdots \cX_k}^{(p)} \equiv \sum_{j_1,\cdots,j_k}
d_{\cX_1\cdots\cX_k}^{j_1\cdots j_k}
\lag \cX_1 \rag_{\r_{j_1}} \cdots \lag \cX_k \rag_{\r_{j_k}}.
\ee

\subsection{Fidelity}

From the replica trick \eqref{replicatrick}, the fidelity \eqref{Fdef} between the RDMs $\r_A$ and $\s_A$ can be obtained as \cite{Sui:2025qve}
\be \label{fidelity}
F(\r_A,\s_A) = 1 + \sum_{k=2}^{+\inf}
\sum_{\{\cX_1,\cdots,\cX_k\}}
\ell^{\D_{\cX_1}+\cdots+\D_{\cX_k}}
\cF_{\cX_1 \cdots \cX_k}^{(1/2)}.
\ee
Note that we have taken the $p\to\f12$ limit of \eqref{tdtrArAsAp}, and that the sum over the number $k$ of quasiprimary operators starts from $k=2$. The single-operator contributions vanish in the $p\to\f12$ limit, so that the fidelity is expanded directly in powers of the differences of the one-point functions.

Some universal contributions $\cF^{(1/2)}_{\cX_1\cdots \cX_k}$ have been obtained in \cite{Sui:2025qve}, and we collect the ones that will be useful in this paper. For a quasiprimary operator $\cX$ with normalization $\a_\cX$, structure constant $C_{\cX\cX\cX}$, scaling dimension $\D_\cX$ and spin $s_\cX$, the contributions of two and three identical operators are
\be
\cF^{(1/2)}_{\cX\cX}
 = -\frac{\G\big(\D_\cX+\f12\big)}{2^{2\D_\cX+3}\sqrt\pi\,\G(\D_\cX+1)}
   \frac{\ii^{2s_\cX}(\lag \cX\rag_\rho-\lag \cX\rag_\sigma)^2}{\a_\cX},
\ee
\be
\cF^{(1/2)}_{\cX\cX\cX}
 = \frac{\G\big(\D_\cX+\f12\big)^2}{2^{2\D_\cX+4}\pi\,\G\big(\f{\D_\cX}2+1\big)\G\big(\f{3\D_\cX}2+1\big)}
   \frac{\ii^{s_\cX}C_{\cX\cX\cX}(\lag \cX\rag_\rho-\lag \cX\rag_\sigma)^2(\lag \cX\rag_\rho+\lag \cX\rag_\sigma)}{\a_\cX^3}.
\ee
For the stress tensor $T$ and a quasiprimary operator $\cX$ with conformal weight $h_\cX$, there is
\bea
&& \cF_{T\cX\cX}^{(1/2)} = \f{h_\cX\G(\D_\cX+\f12)}{2^{2\D_\cX+5} \sqrt{\pi} c \G(\D_\cX+2)}
\f{\ii^{2s_\cX}}{\a_\cX}
(\lag \cX \rag_\r-\lag \cX \rag_\s)
\big[
( \lag T \rag_\r + \lag T \rag_\s ) ( \lag \cX \rag_\r - \lag \cX \rag_\s ) \nn\\
&& \phantom{\cF_{T\cX\cX}^{(1/2)} =}
- \D_\cX (
5 \lag T \rag_\r \lag \cX \rag_\r
+ \lag T \rag_\r \lag \cX \rag_\s
- \lag T \rag_\s \lag \cX \rag_\r
-5 \lag T \rag_\s \lag \cX \rag_\s
)
\big].
\eea
We also need the contribution of four stress tensors, for which we find
\bea
&& \cF^{(1/2)}_{TTTT}
 = -\frac{(\lag T\rag_\rho-\lag T\rag_\sigma)^2\big(203\lag T\rag_\rho^2+298\lag T\rag_\rho\lag T\rag_\sigma+203\lag T\rag_\sigma^2\big)}{524288\,c^2} \nn\\
&& \phantom{F^{(1/2)}_{TTTT}=}
   -\frac{(\lag T\rag_\rho-\lag T\rag_\sigma)^2\big(53\lag T\rag_\rho^2+70\lag T\rag_\rho\lag T\rag_\sigma+53\lag T\rag_\sigma^2\big)}{16384\,c^3}.
\eea

Besides the above known cases, we also need the contribution of two identical quasiprimary operators $\cX$ together with a third, different one $\cY$, in the case that the scaling dimensions and spins are related as $\D_\cY=2\D_\cX$ and $s_\cY=2s_\cX$. A case in point is the pair $(\cX,\cY)=(T,\cA)$ used below.
Summing over the replica indices and performing the analytic continuation $p\to\f12$, we find the universal result
\bea \label{F12XXY}
&& \cF^{(1/2)}_{\cX\cX\cY}
 = \frac{C_{\cX\cX\cY}(\lag \cX\rag_\rho-\lag \cX\rag_\sigma)}{2^{4+8\D_\cX}\a_\cX^2\a_\cY}
   \bigg[(\lag \cX\rag_\rho-\lag \cX\rag_\sigma)(\lag \cY\rag_\rho+\lag \cY\rag_\sigma)\binom{2\D_\cX}{\D_\cX}^2 \nn\\
&& \phantom{F^{(1/2)}_{\cX\cX\cY}=}
   +2(\lag \cX\rag_\rho+\lag \cX\rag_\sigma)(\lag \cY\rag_\rho-\lag \cY\rag_\sigma)\binom{4\D_\cX}{2\D_\cX}\bigg],
\eea
where $C_{\cX\cX\cY}$ is the structure constant and $\a_\cX$, $\a_\cY$ are the normalization constants. The derivation of \eqref{F12XXY} is given in appendix~\ref{appF12XXY}.

\subsection{R\'enyi fidelity}

Compared with the fidelity, or equivalently the BD, the calculation of the R\'enyi fidelity, or equivalently the RBD, is more involved. We calculate the R\'enyi fidelity \eqref{RFdef} between the RDMs $\r_A$ and $\s_A$ from the OPE of twist operators. Using the definition of the normalized trace \eqref{tdtr}, the subsystem R\'enyi fidelity is written as
\be
F_p(\rho_A,\sigma_A) = \frac{\tdtr_A[(\rho_A\sigma_A)^p]}{\sqrt{\tdtr_A(\rho_A^{2p})\tdtr_A(\sigma_A^{2p})}},
\ee
for general integers $p=1,2,\cdots$. The moments $\tdtr_A(\rho_A^{2p})$ and $\tdtr_A(\sigma_A^{2p})$ can be obtained from \eqref{tdtrArAn}. The new ingredient is $\tdtr_A[(\rho_A\sigma_A)^p]$, which is given by \eqref{tdtrArAsAp}, and what remains is to perform the sums over the replica indices in \eqref{FpX1dddXk} systematically.

The simplest case is
\be
\cF^{(p)}_{\cX} = b_\cX^e \lag \cX \rag_\r + b_\cX^o \lag \cX \rag_\s
                = b_\cX^e ( \lag \cX \rag_\r + \lag \cX \rag_\s ),
\ee
with $b_\cX^e=b_\cX^o=p d_\cX$. The superscripts $e$ and $o$ indicate that the corresponding replica indices are even or odd, respectively.

We then consider the example
\be
\cF^{(p)}_{\cX\cX} = \sum_{j_1<j_2} d_{\cX\cX}^{j_1j_2} \lag \cX \rag_{\r_{j_1}} \lag \cX \rag_{\r_{j_2}}.
\ee
We define the coefficients
\be
b_{\cX\cX}^{ee} = \sum_{a_1<a_2} d_{\cX\cX}^{a_1a_2}, ~~
b_{\cX\cX}^{eo} = \sum_{a,b} d_{\cX\cX}^{ab}, ~~
b_{\cX\cX}^{oo} = \sum_{b_1<b_2} d_{\cX\cX}^{b_1b_2},
\ee
satisfying $b_{\cX\cX}^{ee}=b_{\cX\cX}^{oo}$ and $b_{\cX\cX}=2b_{\cX\cX}^{ee}+b_{\cX\cX}^{eo}$, with $b_{\cX\cX}$ being the coefficient entering the moment \eqref{tdtrArAn}. We finally get
\be
\cF^{(p)}_{\cX\cX} = b_{\cX\cX}^{ee} (\lag \cX \rag_\r^2 + \lag \cX \rag_\s^2) + b_{\cX\cX}^{eo} \lag \cX \rag_\r \lag \cX \rag_\s.
\ee

Another example is
\be
\cF^{(p)}_{\cX\cY} = \sum_{j_1 \neq j_2} d_{\cX\cY}^{j_1j_2} \lag \cX \rag_{\r_{j_1}} \lag \cY \rag_{\r_{j_2}}.
\ee
We define the coefficients
\be
b_{\cX\cY}^{ee} = \sum_{a_1 \neq a_2} d_{\cX\cY}^{a_1a_2}, ~~
b_{\cX\cY}^{eo} = \sum_{a,b} d_{\cX\cY}^{ab}, ~~
b_{\cX\cY}^{oe} = \sum_{b,a} d_{\cX\cY}^{ba}, ~~
b_{\cX\cY}^{oo} = \sum_{b_1 \neq b_2} d_{\cX\cY}^{b_1b_2},
\ee
satisfying $b_{\cX\cY}^{ee}=b_{\cX\cY}^{oo}$, $b_{\cX\cY}^{eo}=b_{\cX\cY}^{oe}$, and $b_{\cX\cY}=2(b_{\cX\cY}^{ee}+b_{\cX\cY}^{eo})$. We finally get
\be
\cF^{(p)}_{\cX\cY} = b_{\cX\cY}^{ee} ( \lag \cX \rag_\r \lag \cY \rag_\r + \lag \cX \rag_\s \lag \cY \rag_\s ) + b_{\cX\cY}^{eo} ( \lag \cX \rag_\r \lag \cY \rag_\s + \lag \cX \rag_\s \lag \cY \rag_\r ).
\ee

Coefficients involving more quasiprimary operators can be defined similarly, and we will not list all the forms here. In \cite{Wang:2026coe}, we collect all the coefficients we have used in this paper. With these coefficients, we can calculate the forms
\bea
&& \cF^{(p)}_{\cX\cX\cX} = b_{\cX\cX\cX}^{eee} (\lag \cX \rag_\r^3 + \lag \cX \rag_\s^3)
                         + b_{\cX\cX\cX}^{eeo} \lag \cX \rag_\r \lag \cX \rag_\s ( \lag \cX \rag_\r + \lag \cX \rag_\s ), \nn\\
&& \cF^{(p)}_{\cX\cX\cY} = b_{\cX\cX\cY}^{eee} ( \lag\cX\rag^2_\r\lag\cY\rag_\r + \lag\cX\rag^2_\s\lag\cY\rag_\s )
                         + b_{\cX\cX\cY}^{eeo} ( \lag\cX\rag^2_\r\lag\cY\rag_\s + \lag\cX\rag^2_\s\lag\cY\rag_\r ) \nn\\
&& \phantom{\cF^{(p)}_{\cX\cX\cY} =}
                         + b_{\cX\cX\cY}^{eoe} \lag\cX\rag_\r \lag\cX\rag_\s( \lag\cY\rag_\r + \lag\cY\rag_\s ), \\
&& \cF^{(p)}_{\cX\cX\cX\cX} = b_{\cX\cX\cX\cX}^{eeee} (\lag \cX \rag_\r^4 + \lag \cX \rag_\s^4)
                            + b_{\cX\cX\cX\cX}^{eeeo} \lag \cX \rag_\r \lag \cX \rag_\s ( \lag \cX \rag_\r^2 + \lag \cX \rag_\s^2 )
                            + b_{\cX\cX\cX\cX}^{eeoo} \lag \cX \rag_\r^2 \lag \cX \rag_\s^2. \nn
\eea

\section{Free massless fermion theory} \label{secFermion}

We now consider the 2D free massless fermion theory, which has central charge $c=\f12$, and take the two states to be the ground state $|G\rag$, with conformal weights $(0,0)$, and the state $|\varepsilon\rag$ created by the energy operator $\varepsilon$, with conformal weights $(\f12,\f12)$. For all the states considered in this paper, the contributions from the holomorphic and anti-holomorphic sectors decouple, and so we only need to consider the holomorphic sector.

For the states $|G\rag$ and $|\ve\rag$ in the 2D free massless fermion theory, we need the quasiprimary operators listed in table~\ref{QPs1}; the explicit forms and properties of these operators, as well as their expectation values in the ground state $|G\rag$, can be found in \cite{Chen:2013kpa,Chen:2013dxa}. The expectation values in the excited state $|\varepsilon\rag$ follow directly from the general primary-state results in \cite{Lin:2016dxa,He:2017vyf}.

\begin{table}[t]
  \centering
  \begin{tabular}{r|c|c|c|c}
    \hline
    conformal weight      & 2     & 4       & 6             & 8 \\ \hline
    quasiprimary operator & \(T\) & \(\cA\) & \(\cB,\ \cD\) & \(\cE,\ \cH,\ \cI\) \\ \hline
  \end{tabular}
  \caption{The quasiprimary operators we use in the vacuum conformal family of a single copy of the 2D free massless fermion theory or of a 2D holographic CFT.}\label{QPs1}
\end{table}

From the result of \cite{Zhang:2019itb}, it is easy to get the holomorphic contribution to the fidelity
\be \label{fer0}
F(\r_{A,G},\r_{A,\ve})|_\holo =
\f{\G(\f{3+\csc\f{\pi x}{2}}{4})}
{\G(\f{1+\csc\f{\pi x}{2}}{4})}
\sqrt{2\sin(\pi x)}.
\ee
Using
\be \label{ferFexp}
F(\rho_{A,G},\rho_{A,\ve}) |_\holo
 = 1+\ell^4 F^{(1/2)}_{TT}+\ell^6 F^{(1/2)}_{TTT}
   +\ell^8\big(F^{(1/2)}_{\cA\cA}+F^{(1/2)}_{TT\cA}+F^{(1/2)}_{TTTT}\big)+O(\ell^{10}),
\ee
we obtain the short interval expansion of the fidelity to order $\ell^8$, in agreement with the direct expansion of the exact result \eqref{fer0}.

From the well-known ground-state R\'enyi entropy \cite{Holzhey:1994we,Calabrese:2004eu} and the definition of the normalized trace \eqref{tdtr}, it is easy to get the holomorphic part of the normalized trace
\be \label{fer1}
\tdtr_A \r_{A,G}^n |_\holo = \Big( \f{L}{\pi\ell} \sin \f{\pi\ell}{L} \Big)^{-2h_n}.
\ee
From the R\'enyi entropy of the excited state $|\varepsilon\rag$ \cite{Essler:2012pni,Calabrese:2014ntv}, one can easily get
\be \label{fer2}
\tdtr_A \r_{A,\ve}^n |_\holo = \Big( \f{L}{\pi\ell} \sin \f{\pi\ell}{L} \Big)^{-2h_n}
                               \Big( \f{2}{n} \sin\f{\pi\ell}{L} \Big)^{n}
                               \frac{\G\big(\frac{1+n+n\csc\f{\pi\ell}{L}}{2}\big)}
                                    {\G\big(\frac{1-n+n\csc\f{\pi\ell}{L}}{2}\big)},
\ee
From the result of \cite{Zhang:2026jef}, one gets
\be \label{fer3}
\tdtr_A [ ( \r_{A,G} \r_{A,\ve} )^p ] |_\holo = \Big( \f{L}{\pi\ell} \sin \f{\pi\ell}{L} \Big)^{-2h_{2p}}
                                                \Big( \f{1}{p} \sin\f{\pi\ell}{L} \Big)^{p}
                                                \frac{\G\big(\frac{1+p+p\csc\f{\pi\ell}{2L}}{2}\big)}
                                                     {\G\big(\frac{1-p+p\csc\f{\pi\ell}{2L}}{2}\big)}.
\ee
In CFT$^n$ we need the quasiprimary operators listed in table~\ref{CFTnQPs1}, and with these we reproduce the short interval expansions of the results \eqref{fer1}, \eqref{fer2} and \eqref{fer3} to order $\ell^8$.

\begin{table}[t]
  \centering
  \begin{tabular}{c|l}
    \hline
    conformal weight & quasiprimary operators \\ \hline
    2 & \(T\) \\ \hline
    4 & \(\cA,\ TT\) \\ \hline
    6 & \(\cB,\ \cD,\ T\cA,\ TTT\) \\ \hline
    8 & \(\cE,\ \cH,\ \cI,\ T\cB,\ T\cD,\ \cA\cA,\ TT\cA,\ TTTT\) \\ \hline
  \end{tabular}
  \caption{The CFT$^n$ quasiprimary operators we use in this paper to calculate the short interval expansion of the R\'enyi entropy and R\'enyi fidelity in the 2D free massless fermion theory and in holographic CFTs.}\label{CFTnQPs1}
\end{table}

\section{Free massless boson theory} \label{secBoson}

We then turn to the 2D free massless boson theory, which has central charge $c=1$, and consider the ground state $|G\rag$, with conformal weights $(0,0)$, and the current state $|J\rag$, with conformal weights $(1,0)$.

In the free massless boson theory, we need the quasiprimary operators listed in table~\ref{QPs2}. In addition to the operators of table~\ref{QPs1}, there is now the primary operator $\cL$ with conformal weights $(4,0)$, defined as \cite{Zhang:2020txb}
\be \label{primaryL}
\cL \equiv (J(J(JJ))) +\f32(\p J \p J) - (\p^2 J J).
\ee
From the algebra of the current modes, we get the normalization and structure constant
\be
\a_\cL=54, ~~ C_{JJ\cL} = -6.
\ee
In addition to the expectation values collected in \cite{Chen:2013kpa,Chen:2013dxa,Lin:2016dxa,He:2017vyf}, we also need
\be
\lag \cL \rag_G = 0, ~~
\lag \cL \rag_J = -\f{96\pi^4}{L^4}.
\ee

\begin{table}[t]
  \centering
  \begin{tabular}{r|c|c|c|c}
    \hline
    conformal weight      & 2     & 4             & 6             & 8 \\ \hline
    quasiprimary operator & \(T\) & \(\cA,\ \cL\) & \(\cB,\ \cD\) & \(\cE,\ \cH,\ \cI\) \\ \hline
  \end{tabular}
  \caption{The quasiprimary operators we use in the vacuum conformal family of a single copy of the 2D free massless boson theory.}\label{QPs2}
\end{table}

The exact fidelity is \cite{Zhang:2019itb}
\be \label{bos0}
F(\r_{A,G},\r_{A,J}) =
\f{\G^2(\f{3+\csc\f{\pi x}{2}}{4})}
{\G^2(\f{1+\csc\f{\pi x}{2}}{4})}
2\sin(\pi x).
\ee
Only the holomorphic sector contributes to the fidelity $F(\r_{A,G},\r_{A,J})$. Using
\be
F(\rho_{A,G},\rho_{A,J})
 = 1+\ell^4 F^{(1/2)}_{TT}+\ell^6 F^{(1/2)}_{TTT}
   +\ell^8\big(F^{(1/2)}_{\cA\cA}+F^{(1/2)}_{\cL\cL}+F^{(1/2)}_{TT\cA}+F^{(1/2)}_{TTTT}\big)+O(\ell^{10}),
\ee
we obtain the short interval expansion of \eqref{bos0} to order $\ell^8$, again in agreement with the direct expansion of the exact result.

From the R\'enyi entropy of the state $|J\rag$ \cite{Essler:2012pni,Calabrese:2014ntv}, one can easily get
\be \label{bos2}
\tdtr_A \r_{A,J}^n |_\holo = \Big( \f{L}{\pi\ell} \sin \f{\pi\ell}{L} \Big)^{-2h_n}
                               \Big( \f{2}{n} \sin\f{\pi\ell}{L} \Big)^{2n}
                               \frac{\G^2\big(\frac{1+n+n\csc\f{\pi\ell}{L}}{2}\big)}
                                    {\G^2\big(\frac{1-n+n\csc\f{\pi\ell}{L}}{2}\big)},
\ee
From the result of \cite{Zhang:2026jef}, one gets
\be \label{bos3}
\tdtr_A [ ( \r_{A,G} \r_{A,J} )^p ] |_\holo = \Big( \f{L}{\pi\ell} \sin \f{\pi\ell}{L} \Big)^{-2h_{2p}}
                                                \Big( \f{1}{p} \sin\f{\pi\ell}{L} \Big)^{2p}
                                                \frac{\G^2\big(\frac{1+p+p\csc\f{\pi\ell}{2L}}{2}\big)}
                                                     {\G^2\big(\frac{1-p+p\csc\f{\pi\ell}{2L}}{2}\big)}.
\ee
In CFT$^n$ we need the quasiprimary operators listed in table~\ref{CFTnQPs2}, and with these we reproduce the short interval expansions of the results \eqref{bos2} and \eqref{bos3} to order $\ell^8$.

\begin{table}[t]
  \centering
  \begin{tabular}{c|l}
    \hline
    conformal weight & quasiprimary operators \\ \hline
    2 & \(T\) \\ \hline
    4 & \(\cA,\ TT\) \\ \hline
    6 & \(\cB,\ \cD,\ T\cA,\ TTT\) \\ \hline
    8 & \(\cE,\ \cH,\ \cI,\ T\cB,\ T\cD,\ \cA\cA,\ \cL\cL,\ TT\cA,\ TTTT\) \\ \hline
  \end{tabular}
  \caption{The CFT$^n$ quasiprimary operators we use in this paper to calculate the short interval expansion of the R\'enyi entropy and R\'enyi fidelity in the 2D free massless boson theory.}\label{CFTnQPs2}
\end{table}

\section{Holographic CFTs} \label{secHoloCFT}

The AdS$_3$/CFT$_2$ correspondence \cite{Maldacena:1997re} relates three-dimensional gravity with Newton constant $G$ and AdS radius $R$ to a two-dimensional CFT with a large central charge $c=\frac{3R}{2G}$ \cite{Brown:1986nw} and a sparse spectrum of low-lying operators \cite{Hartman:2014oaa}. In this duality a black hole microstate can be described by a high-energy primary state of the CFT \cite{Strominger:1997eq}, more precisely by an energy eigenstate dual to a non-rotating Ba\~nados-Teitelboim-Zanelli (BTZ) black hole \cite{Banados:1992wn}. Whether such eigenstates can be distinguished from a thermal ensemble, or from one another, by observations on a small subsystem is a form of the black hole information problem \cite{Hawking:1976ra,Almheiri:2019psf,Penington:2019npb}, and at the same time a direct test of the eigenstate thermalization hypothesis (ETH) \cite{Deutsch:1991msp,Srednicki:1994mfb,Deutsch:2018thx}. Related questions on the perturbative distinguishability of black hole microstates have been addressed recently in \cite{Zhang:2025vaq,Sui:2025qve}.

We apply the method of section~\ref{secMethod} to such holographic CFTs. We consider two types of states, namely a heavy primary state $|\phi\rag$ with conformal weight $h$ on a circle of length $L$ and the thermal state $\r_\b$ with inverse temperature $\b$ on an infinite line. We take a subsystem $A=[0,\ell]$, keep only the contributions of the vacuum conformal family, and use the fact that the holomorphic and anti-holomorphic sectors factorize, so that it is enough to discuss the holomorphic sector. The quasiprimary operators of a single copy of the CFT are listed in table~\ref{QPs1}; their expectation values in primary states can be found in \cite{Lin:2016dxa,He:2017vyf}, and those in the thermal state in \cite{Chen:2013kpa,Chen:2013dxa}.

From the OPE of twist operators, with the replica coefficients collected in \cite{Wang:2026coe}, we calculate the BD $B=-\log F$ \eqref{BD} and the RBD $B_p=-\log F_p$ \eqref{RBD} of two states. Taking the logarithm is also what makes the large $c$ expansion transparent. In the short interval expansion of the fidelity and the R\'enyi fidelity, the term of order $\ell^8$ contains a piece of order $c^2$ that is nothing but the square of the order $\ell^4$ term, and it is precisely cancelled by the expansion of the logarithm, so that the surviving coefficients measure the genuine distinguishability of the two states. Since $c=\frac{3R}{2G}$, a contribution of order $c$ is the classical answer in the bulk, whereas a contribution of order $1/c$ is suppressed by one power of $G$ and is therefore a perturbative quantum gravity correction. This $c$-counting is used throughout this section to decide whether two states can be told apart classically, or only by going beyond classical gravity.

\subsection{Two heavy primary states}

Let $|\phi_1\rangle$ and $|\phi_2\rangle$ be two heavy primary states with conformal weights $h_i=c\varepsilon_i$, where $\varepsilon_i\sim O(c^0)$. These are the states dual to black hole microstates, and, being exact energy eigenstates, they avoid the ambiguities of the fixed-area states of the bulk, which are superpositions of eigenstates \cite{Kudler-Flam:2021rpr,Kudler-Flam:2021alo}. Expanding at large $c$ with $\varepsilon_i$ held fixed, we get the short interval expansion of the BD
\begin{equation} \label{Bph1phi2}
\begin{split}
B(\rho_{A,\phi_1},\rho_{A,\phi_2})
 ={}& \Big[ \frac{3\pi^4c(\varepsilon_1-\varepsilon_2)^2}{32}+O(c^0) \Big]  \Big(\frac{\ell}{L}\Big)^4\\
 &+ \Big[ \frac{\pi^6c(\varepsilon_1-\varepsilon_2)^2\big[1-12(\varepsilon_1+\varepsilon_2)\big]}{64}
     +O(c^0) \Big] \Big(\frac{\ell}{L}\Big)^6 \\
 &+\Big[ \frac{\pi^8c(\varepsilon_1-\varepsilon_2)^2}{5120}
     \big[11-300(\varepsilon_1+\varepsilon_2)\\
 &\qquad\qquad +2700(\varepsilon_1^2+\varepsilon_2^2)+2520\varepsilon_1\varepsilon_2\big]
     +O(c^0) \Big] \Big(\frac{\ell}{L}\Big)^8
  +O(\ell^{10}).
\end{split}
\end{equation}
The leading term is of order $c$: two black hole microstates whose energies differ by an amount of order $c$ are distinguishable already at the classical level in the bulk, as the dual BTZ black holes have different masses and hence different classical metrics.
The more interesting situation is that of two microstates with the same energy at leading order, $h_i=c\varepsilon+\delta_i$ with $\delta_i\sim O(c^0)$, so that the two energies differ only at order $c^0$. Then the order $c$ terms cancel and one gets
\begin{equation}
\begin{split}
B(\rho_{A,\phi_1},\rho_{A,\phi_2})
 ={}& \Big[\frac{3\pi^4(\delta_1-\delta_2)^2}{32c}+O(c^{-2})\Big]
     \Big(\frac{\ell}{L}\Big)^4\\
 &+\Big[\frac{\pi^6(\delta_1-\delta_2)^2(1-24\varepsilon)}{64c}+O(c^{-2})\Big]
     \Big(\frac{\ell}{L}\Big)^6\\
 &+\Big[\frac{\pi^8(\delta_1-\delta_2)^2\big[11+120\varepsilon(66\varepsilon-5)\big]}
        {5120c}+O(c^{-2})\Big]
     \Big(\frac{\ell}{L}\Big)^8
  +O(\ell^{10}).
\end{split}
\end{equation}
Now the distance is of order $1/c$ at every order in $\ell$. The two microstates are indistinguishable in the classical bulk, and it is only the perturbative $1/c$ quantum gravity correction that makes them distinguishable. In the bulk the two BTZ black holes have the same mass at leading order, and the difference between them is a quantum gravity effect. Note that the order $\ell^8$ coefficient is nonzero, so that this perturbative distinguishability is not an artefact of the low orders but persists at the highest order we have computed.

The short interval expansion of the RBD, again at large $c$ with $\varepsilon_i$ held fixed, is
\begin{equation}
\begin{split}
B_p(\rho_{A,\phi_1},\rho_{A,\phi_2})
 ={}& \Big[\frac{(2+p^2)\pi^4c(\varepsilon_1-\varepsilon_2)^2}{48p}+O(c^0)\Big]
     \Big(\frac{\ell}{L}\Big)^4\\
 &+\Big[\frac{\pi^6c(\varepsilon_1-\varepsilon_2)^2
        \big[4+5p^2+6p^4+8(p^2-1)(11+p^2)(\varepsilon_1+\varepsilon_2)\big]}
       {2880p^3}+O(c^0)\Big]
     \Big(\frac{\ell}{L}\Big)^6\\
 &+\Big[\frac{\pi^8c(\varepsilon_1-\varepsilon_2)^2}{2419200\,p^5}
    \big\{4\big[27-1418(\varepsilon_1+\varepsilon_2)
        +18428(\varepsilon_1^2+\varepsilon_2^2)+21816\varepsilon_1\varepsilon_2\big]\\
 &\qquad\qquad +42p^2\big[3+88(\varepsilon_1+\varepsilon_2)
        -3388(\varepsilon_1^2+\varepsilon_2^2)-5016\varepsilon_1\varepsilon_2\big]\\
 &\qquad\qquad +84p^4\big[3+58(\varepsilon_1+\varepsilon_2)
        +332(\varepsilon_1^2+\varepsilon_2^2)+504\varepsilon_1\varepsilon_2\big]\\
 &\qquad\qquad +p^6\big[459+884(\varepsilon_1+\varepsilon_2)
        -884(\varepsilon_1^2+\varepsilon_2^2)-2088\varepsilon_1\varepsilon_2\big]\big\}
    +O(c^0)\Big]\Big(\frac{\ell}{L}\Big)^8\\
 &+O(\ell^{10}).
\end{split}
\end{equation}
At $p=\frac12$ this reduces to \eqref{Bph1phi2}, as required by $B=B_{1/2}$.
For $h_i=c\varepsilon+\delta_i$ with $\delta_i\sim O(c^0)$ one gets instead
\begin{equation}
\begin{split}
B_p(\rho_{A,\phi_1},\rho_{A,\phi_2})
 ={}& \Big[\frac{(2+p^2)\pi^4(\delta_1-\delta_2)^2}{48\,c\,p}+O(c^{-2})\Big]
     \Big(\frac{\ell}{L}\Big)^4\\
 &+\Big[\frac{\pi^6(\delta_1-\delta_2)^2
        \big[4-176\varepsilon+2p^4(3+8\varepsilon)+5p^2(1+32\varepsilon)\big]}
       {2880\,c\,p^3}+O(c^{-2})\Big]
     \Big(\frac{\ell}{L}\Big)^6\\
 &+\Big[\frac{\pi^8(\delta_1-\delta_2)^2}{2419200\,c\,p^5}
    \big\{p^6\big[459+8(221-482\varepsilon)\varepsilon\big]
    +42p^2\big[3+176(1-67\varepsilon)\varepsilon\big]\\
 &\qquad\qquad +84p^4\big[3+4\varepsilon(29+292\varepsilon)\big]\\
 &\qquad\qquad +4\big[27+4\varepsilon(14668\varepsilon-709)\big]\big\}
    +O(c^{-2})\Big]\Big(\frac{\ell}{L}\Big)^8
  +O(\ell^{10}).
\end{split}
\end{equation}
which is again of order $1/c$ at every order in $\ell$, so that the classical-versus-perturbative conclusion is not specific to the BD but holds for the whole one-parameter family of RBDs.

\subsection{Two thermal states}

We then consider two thermal states $\rho_{\beta_1}$ and $\rho_{\beta_2}$ with inverse temperatures $\beta_1$ and $\beta_2$ on the infinite line, a pair of states that has not been considered before. The short interval expansion of the BD is
\begin{equation}
\begin{split}
B(\rho_{A,\beta_1},\rho_{A,\beta_2})
 ={}& \frac{\pi^4c(\beta_1^2-\beta_2^2)^2}{6144\,\beta_1^4\beta_2^4}\,\ell^4
  -\frac{\pi^6c(\beta_1^2-\beta_2^2)^2(\beta_1^2+\beta_2^2)}
        {73728\,\beta_1^6\beta_2^6}\,\ell^6\\
 &+\frac{\pi^8c(\beta_1^2-\beta_2^2)^2(53\beta_1^4+70\beta_1^2\beta_2^2+53\beta_2^4)}
        {47185920\,\beta_1^8\beta_2^8}\,\ell^8
  +O(\ell^{10}).
\end{split}
\end{equation}
For two temperatures that differ by an amount of order $c^0$, this is of order $c$, and the two thermal states are classically distinct: they are dual to black holes of different temperature, hence to different classical geometries. If instead the two inverse temperatures coincide at leading order, $\beta_i=\beta+\frac{\gamma_i}{c}$, one gets
\begin{equation}
\begin{split}
B(\rho_{A,\beta_1},\rho_{A,\beta_2})
 ={}& \Big[\frac{\pi^4(\gamma_1-\gamma_2)^2}{1536\,\beta^6c}+O(c^{-2})\Big]\,\ell^4
  -\Big[\frac{\pi^6(\gamma_1-\gamma_2)^2}{9216\,\beta^8c}+O(c^{-2})\Big]\,\ell^6\\
 &+\Big[\frac{11\pi^8(\gamma_1-\gamma_2)^2}{737280\,\beta^{10}c}+O(c^{-2})\Big]\,\ell^8
  +O(\ell^{10}).
\end{split}
\end{equation}
which is again of order $1/c$. The structure is completely parallel to that of the two heavy primary states in the previous subsection. Two thermal states whose temperatures agree to leading order can only be distinguished by the $1/c$ correction, and the conclusion holds up to order $\ell^8$.

The short interval expansion of the RBD is
\begin{equation}
\begin{split}
B_p(\rho_{A,\beta_1},\rho_{A,\beta_2})
 ={}& \frac{(2+p^2)\pi^4c(\beta_1^2-\beta_2^2)^2}{27648\,p\,\beta_1^4\beta_2^4}\,\ell^4
  -\frac{(19+14p^2)\pi^6c(\beta_1^2-\beta_2^2)^2(\beta_1^2+\beta_2^2)}
        {3317760\,p\,\beta_1^6\beta_2^6}\,\ell^6\\
 &+\frac{\pi^8c(\beta_1^2-\beta_2^2)^2}
        {7431782400\,p\,\beta_1^8\beta_2^8}
    \big[(3412+3047p^2)(\beta_1^4+\beta_2^4)\\
 &\qquad\qquad\qquad\qquad +2(2252+2017p^2)\beta_1^2\beta_2^2\big]\,\ell^8
  +O(\ell^{10}).
\end{split}
\end{equation}
For $\beta_i=\beta+\frac{\gamma_i}{c}$ one gets
\begin{equation}
\begin{split}
B_p(\rho_{A,\beta_1},\rho_{A,\beta_2})
 ={}& \Big[\frac{(2+p^2)\pi^4(\gamma_1-\gamma_2)^2}{6912\,p\,\beta^6c}+O(c^{-2})\Big]\,\ell^4
  -\Big[\frac{(19+14p^2)\pi^6(\gamma_1-\gamma_2)^2}{414720\,p\,\beta^8c}+O(c^{-2})\Big]\,\ell^6\\
 &+\Big[\frac{(236+211p^2)\pi^8(\gamma_1-\gamma_2)^2}
        {38707200\,p\,\beta^{10}c}+O(c^{-2})\Big]\,\ell^8
  +O(\ell^{10}).
\end{split}
\end{equation}
which is again of order $1/c$ at every order in $\ell$, as for the BD.

\subsection{A heavy primary state and a thermal state}

We finally compare a heavy primary state $|\phi\rangle$ of conformal weight $h$ on the circle with the thermal state $\rho_\beta$ on the infinite line to which it is related by the thermal state hypothesis. The hypothesis states that, as far as the local data are concerned, the heavy state is well approximated by a thermal state when \cite{Fitzpatrick:2014vua,Fitzpatrick:2015zha}
\begin{equation} \label{betaL}
\beta = \frac{L}{\sqrt{24h/c-1}}\,,
\end{equation}
For this value of $\beta$ the two states have the same one-point function of the stress tensor, $\langle T\rangle_\phi=\langle T\rangle_\beta$, and the same energy. Because of this, all the contributions that are proportional to $\langle T\rangle_\phi-\langle T\rangle_\beta$ vanish, and the order $\ell^2$, $\ell^4$ and $\ell^6$ terms of the expansion are absent. The first nonvanishing contribution comes at order $\ell^8$, from the operator $\mathcal{A}$, and we find
\begin{equation}
\begin{split}
B(\rho_{A,\phi},\rho_{A,\beta})
 ={}& \frac{7\pi^8c(11L^4+10L^2\beta^2-\beta^4)^2}
        {42467328\,(5c+22)\,\beta^8}\Big(\frac{\ell}{L}\Big)^8
  +O(\ell^{10}).
\end{split}
\end{equation}
At large $c$ this becomes
\begin{equation}
\begin{split}
B(\rho_{A,\phi},\rho_{A,\beta})
 ={}& \Big[\frac{7\pi^8(11L^4+10L^2\beta^2-\beta^4)^2}
        {212336640\,\beta^8}+O(c^{-1})\Big]\Big(\frac{\ell}{L}\Big)^8
  +O(\ell^{10}).
\end{split}
\end{equation}
which is of order $c^0$, that is, of order $1/c$ relative to the classical answer: a heavy state and the thermal state to which it is dual at the level of $\langle T\rangle$ are again distinguished only by perturbative quantum gravity corrections. In terms of $\varepsilon\equiv h/c$ and using the thermal state hypothesis \eqref{betaL}, the same result can be rewritten as $\frac{7\pi^8c\,\varepsilon^2(22\varepsilon-1)^2}{512(5c+22)}\big(\frac{\ell}{L}\big)^8$, and in the limit $L\to\infty$ at fixed $\beta$ it reduces to $\frac{847\pi^8\ell^8}{212336640\,\beta^8}$.

It is worth emphasizing that this $1/c$ suppression is specific to the BD. Keeping $p$ general, the short interval expansion of the RBD is
\begin{equation} \label{qrbdpb}
\begin{split}
B_p(\rho_{A,\phi},\rho_{A,\beta})
 ={}& \frac{\pi^8c}{100329062400\,(5c+22)\,p^5\beta^8}
    \big[740-2016p^2+3864p^4+2515p^6\\
 &\qquad\qquad\qquad\qquad +35c(1-4p^2)^2(2+p^2)\big]\\
 &\qquad\qquad\qquad\qquad\times(-11L^4-10L^2\beta^2+\beta^4)^2
    \Big(\frac{\ell}{L}\Big)^8
  +O(\ell^{10}).
\end{split}
\end{equation}
and its large-$c$ limit reads
\begin{equation} \label{qrbdpbc}
\begin{split}
B_p(\rho_{A,\phi},\rho_{A,\beta})
 ={}&\frac{\pi^8(11L^4+10L^2\beta^2-\beta^4)^2}{p^5\beta^8}
    \Big[\frac{c(1-4p^2)^2(2+p^2)}{14332723200}
          +\frac{144+98p^2+56p^4+17p^6}{167215104000} \\
 &\qquad\qquad\qquad +O(c^{-1})\Big]\Big(\frac{\ell}{L}\Big)^8
  +O(\ell^{10}).
\end{split}
\end{equation}
The order-$c$ part is proportional to $(1-4p^2)^2$, so it cancels only at $p=\frac12$; for a generic RBD the two states are distinguishable already at order $c$, and it is only the BD, that is, the $p\to\frac12$ limit, that exhibits the perturbative $1/c$ distinguishability.

The different $c$ orders of the BD and the RBD are reminiscent of the different $c$ orders of the entanglement entropy and the R\'enyi entropy. The states $\rho_{A,\phi}$ and $\rho_{A,\beta}$ with $\beta$ given by \eqref{betaL} have the same short interval order-$c$ entanglement entropy \cite{Fitzpatrick:2014vua,Fitzpatrick:2015zha} but different order-$c$ R\'enyi entropies \cite{Lashkari:2016vgj,Lin:2016dxa,He:2017vyf}.

\section{Conclusion} \label{secConc}

In this paper we have studied the subsystem quantum Bhattacharyya distance and quantum R\'enyi-Bhattacharyya distance in 2D CFTs. Extending the twist-operator OPE method of \cite{Sui:2025qve} from the fidelity to the R\'enyi fidelity, we have computed the short interval expansions of the BD and RBD to order $\ell^8$. The method has been validated in the free massless fermion and boson theories, where it reproduces the short interval expansions of the exact results.

Applied to holographic CFTs, the method reveals a uniform structure. Whenever the two states differ in their classical, order-$c$ data, such as the energies of two heavy primary states or the temperatures of two thermal states, the distances are of order $c$, and the states are distinguishable already in classical gravity. When the leading data coincide, the distances are of order $1/c$ at every order of the short interval expansion, and the states can only be distinguished by perturbative quantum gravity effects. For a heavy primary state and the thermal state with the same energy, the BD starts only at order $\ell^8$ and is of order $c^0$, whereas the RBD at a generic R\'enyi index is of order $c$, and the classical contribution cancels only in the fidelity limit $p\to\f12$. This is consistent with the fact that the two states share the same leading entanglement entropy but have different leading R\'enyi entropies.

There are several directions worth exploring. It would be interesting to go beyond the vacuum conformal family and include the non-vacuum contributions \cite{Zhang:2016rja}, to extend the analysis to descendant states \cite{Guo:2018pvi,Brehm:2020zri}, and to clarify the role of the higher conserved KdV charges, which in 2D CFTs at large central charge replace the thermal ensemble by a generalized Gibbs ensemble \cite{Dymarsky:2018lhf,Dymarsky:2019etq,Chen:2024lji,Chen:2024ysb}.

\appendix

\section*{Acknowledgements}

This work is supported by the National Natural Science Foundation of China (Grant No. 12205217) and the Tianjin University Self-Innovation Fund Extreme Basic Research Project (Grant No. 2025XJ21-0007).

\section{Derivation of the universal contribution $F^{(1/2)}_{\cX\cX\cY}$} \label{appF12XXY}

We consider the quasiprimary operators $\cX$ and $\cY$ in a general 2D CFT, requiring that $\D_\cY=2\D_\cX$ and $s_\cY=2s_\cX$. In CFT$^n$ with even integer $n=2p$ there is the quasiprimary operator
\be
\cX_{j_1}\cX_{j_2}\cY_{j_3}.
\ee
For the replica indices $j_1,j_2,j_3\in\{0,1,\cdots,2p-1\}$, there are the constraints $j_1<j_2$, $j_1 \neq j_3$ and $j_2 \neq j_3$. Using the formula of \cite{Calabrese:2010he}, we get the twist operator OPE coefficient
\be
d_{\cX\cX\cY}^{j_1j_2j_3} = \f{C_{\cX\cX\cY}}{(2n)^{4\D_\cX}\a_\cX^2\a_\cY} \f{1}{|s_{j_1j_3}s_{j_2j_3}|^{2\D_\cX}} + O(n-1),
\ee
where $C_{\cX\cX\cY}$ is the structure constant, $\a_\cX$ and $\a_\cY$ are normalization constants of $\cX$ and $\cY$, and we use the shorthand $s_{j_1j_3}\equiv\sin{\frac{\pi(j_1-j_3)}{n}}$ and $s_{j_2j_3}\equiv\sin{\frac{\pi(j_2-j_3)}{n}}$.

Following \cite{Sui:2025qve}, we denote the even replica indices by $a \in \{0,2,\cdots,2p-2\}$ and the odd replica indices by $b \in \{1,3,\cdots,2p-1\}$. We define
\bea
&& F_1(p,\D) \equiv \sum_{a_1<a_2,a_1 \neq a_3,a_2 \neq a_3} \f{1}{|s_{a_1a_3}s_{a_2a_3}|^{2\D}}, \nn\\
&& F_2(p,\D) \equiv \sum_{a_1<a_2,b} \f{1}{|s_{a_1b}s_{a_2b}|^{2\D}}, \nn\\
&& F_3(p,\D) \equiv \sum_{a_1 \neq a_2,b} \f{1}{|s_{a_1a_2}s_{ba_2}|^{2\D}}.
\eea
Evaluating these sums for several values of $\D$ gives the specific values
\begin{center}
\begin{tabular}{c|cccccc}
$\D$           & 1 & 2 & 3 & 4 & 5 & 6 \\ \hline
$F_1(1/2,\D)$ & $\frac{1}{16}$ & $\frac{11}{256}$ & $\frac{281}{8192}$ & $\frac{1915}{65536}$ & $\frac{54127}{2097152}$ & $\frac{782761}{33554432}$ \\
$F_2(1/2,\D)$ & $-\frac{1}{32}$ & $-\frac{13}{512}$ & $-\frac{181}{8192}$ & $-\frac{2605}{131072}$ & $-\frac{38251}{2097152}$ & $-\frac{569317}{33554432}$ \\
$F_3(1/2,\D)$ & $-\frac{1}{32}$ & $-\frac{9}{512}$ & $-\frac{25}{2048}$ & $-\frac{1225}{131072}$ & $-\frac{3969}{524288}$ & $-\frac{53361}{8388608}$ \\
\end{tabular}
\end{center}
from which one can use Seed-2.1-Pro and obtain the general expressions
\bea
&& F_1(1/2,\D) = \f{1}{2^{4 \Delta +4}} \bigg[ \binom{2 \Delta }{\Delta }^2 + 2 \binom{4 \Delta }{2 \Delta} \bigg], \nn\\
&& F_2(1/2,\D) = \f{1}{2^{4 \Delta +4}} \bigg[ \binom{2 \Delta }{\Delta }^2 - 2 \binom{4 \Delta }{2 \Delta} \bigg], \nn\\
&& F_3(1/2,\D) = - \f{1}{2^{4 \Delta +3}} \binom{2 \Delta }{\Delta }^2.
\eea
It is easy to check
\be
F_1(1/2,\D) + F_2(1/2,\D) + F_3(1/2,\D) = 0.
\ee
Finally, we get the universal contribution to the fidelity
\bea
&& \cF_{\cX\cX\cY}^{(1/2)} = \frac{C_{\cX\cX\cY}}{2^{4\D_\cX}\a_\cX^2\a_\cY}\big[
                          F_1(1/2,\D_\cX) ( \lag\cX\rag^2_\r\lag\cY\rag_\r + \lag\cX\rag^2_\s\lag\cY\rag_\s )  \\
&& \phantom{\cF_{\cX\cX\cY}^{(1/2)} =}
                        + F_2(1/2,\D_\cX) ( \lag\cX\rag^2_\r\lag\cY\rag_\s + \lag\cX\rag^2_\s\lag\cY\rag_\r )
                        + F_3(1/2,\D_\cX) \lag\cX\rag_\r \lag\cX\rag_\s( \lag\cY\rag_\r + \lag\cY\rag_\s )
                        \big],\nn
\eea
which equals \eqref{F12XXY}.


\providecommand{\href}[2]{#2}\begingroup\raggedright\endgroup

\end{document}